\documentclass[conference]{IEEEtran}
\IEEEoverridecommandlockouts
\usepackage{cite}
\usepackage{amsmath,amssymb,amsfonts}
\usepackage{algorithmic}
\usepackage{graphicx}
\usepackage{textcomp}
\usepackage{xcolor}
\usepackage{booktabs} 
\usepackage{amsthm}
\def\BibTeX{{\rm B\kern-.05em{\sc i\kern-.025em b}\kern-.08em
    T\kern-.1667em\lower.7ex\hbox{E}\kern-.125emX}}

\begin{document}

\title{Empowering Digital Agriculture: A Privacy-Preserving Framework for Data Sharing and Collaborative Research
}

\author{\IEEEauthorblockN{Osama Zafar}
\IEEEauthorblockA{\textit{Department of Computer and Data Sciences} \\
\textit{Case Western Reserve University}\\
Cleveland, OH 44106, USA \\
oxz23@case.edu}
\and
\IEEEauthorblockN{Rosemarie Santa González}
\IEEEauthorblockA{\textit{School of Computer Science} \\
\textit{Georgia Institute of Technology}\\
Atlanta, GA 30332, USA\\
rosemarie.santa@gatech.edu}
\and
\IEEEauthorblockN{Mina Namazi}
\IEEEauthorblockA{\textit{Department of Computer and Data Sciences} \\
\textit{Case Western Reserve University}\\
Cleveland, OH 44106, USA \\
mxn559@case.edu}
\and
\IEEEauthorblockN{Alfonso Morales}
\IEEEauthorblockA{\textit{Department of Planning and Landscape Architecture} \\
\textit{University of Wisconsin–Madison}\\
Madison, WI 53706, USA \\
morales1@wisc.edu}
\and
\IEEEauthorblockN{Erman Ayday}
\IEEEauthorblockA{\textit{Department of Computer and Data Sciences} \\
\textit{Case Western Reserve University}\\
Cleveland, OH 44106, USA \\
exa208@case.edu}
}

\maketitle

\begin{abstract}
Data-driven agriculture, which integrates technology and data into agricultural practices, has the potential to improve crop yield, disease resilience, and long-term soil health. However, privacy concerns, such as adverse pricing, discrimination, and resource manipulation, deter farmers from sharing data, as it can be used against them. To address this barrier, we propose a privacy-preserving framework that enables secure data sharing and collaboration for research and development while mitigating privacy risks. The framework combines dimensionality reduction techniques (like Principal Component Analysis (PCA)) and differential privacy by introducing Laplacian noise to protect sensitive information. The proposed framework allows researchers to identify potential collaborators for a target farmer and train personalized machine learning models either on the data of identified collaborators via federated learning or directly on the aggregated privacy-protected data. It also allows farmers to identify potential collaborators based on similarities. We have validated this on real-life datasets, demonstrating robust privacy protection against adversarial attacks and utility performance comparable to a centralized system. We demonstrate how this framework can facilitate collaboration among farmers and help researchers pursue broader research objectives. The adoption of the framework can empower researchers and policymakers to leverage agricultural data responsibly, paving the way for transformative advances in data-driven agriculture. By addressing critical privacy challenges, this work supports secure data integration, fostering innovation and sustainability in agricultural systems.
\end{abstract}


\section{Introduction} 
\label{section:intro}

Agriculture has long been the backbone of human civilization, driving innovation to enhance food production and promote sustainability. Over the years, advances such as fertilizers, pesticides, genetically modified seeds, and soil composition analysis have transformed cultivation and harvest practices \cite{pretty2008agricultural, tilman2002agricultural}. Today, this tradition of innovation is driven by a wealth of data collected from satellites, unmanned aerial vehicles, self-driving machines, and field sensors \cite{wolfert2017big}. Although farmers occasionally share sales data for marketing purposes, sharing sensitive data on environmental and soil conditions presents significant privacy challenges \cite{demets}. The very data that empower precision agriculture to optimize yields and sustainability also carry risks of misuse, such as adverse pricing, price discrimination, disease interference, inflated insurance costs, and harmful competition \cite{wiseman2019farmers, taylor2018climate}. These concerns deter farmers from sharing their data, creating a critical barrier to advancing agricultural research. To unlock the full potential of data-driven agriculture, it is imperative to prioritize protecting farmers from these risks, ensuring that innovation does not come at the expense of their trust and livelihoods.


The existing data collection and collaboration frameworks in the domain, like Farm2Facts \cite{farm2fact,ledesma}, Local Line \cite{localLine}, and Open Food Network \cite{openFoodNetwork}, collect vast agricultural data, but they lack mechanisms for privacy-preserving collaboration. As a result, farmers and researchers are unable to share data freely on such platforms due to proprietary or privacy concerns, which impede innovation. Publicly sharing sensitive agricultural data raises concerns about privacy and security, even for research purposes. Several critical security threats in the digital agriculture domain pose significant risks to farmers' privacy and data security. These threats can be intentional attacks perpetrated by malicious third parties, such as attribute inference attacks and membership inference attacks, or they can be unintentional risks, including configuration errors and improper encryption, which may inadvertently expose sensitive information.

This paper proposes a framework to address these privacy and security challenges in data-driven agriculture. The framework is designed to mitigate the risks of data misuse while fostering trust among farmers. It achieves this by employing techniques such as differential privacy and dimensionality reduction, which obscure sensitive data and prevent unauthorized reconstruction of the original datasets. Furthermore, the framework aims to reduce the risk of membership inference attacks by ensuring that even anonymized data remains untraceable to individuals. By addressing these specific threats, the framework creates a secure environment for data sharing and collaboration for the farmers, thereby promoting trust and enabling the agricultural research community to fully leverage the potential of data-driven agriculture.

The main goal of this work is the application of privacy-preserving machine learning tools like dimensionality reduction, differential privacy, etc, to address a critical privacy challenge in digital agriculture: enabling collaborative research while safeguarding data privacy. Our research makes a significant contribution by developing a comprehensive privacy-preserving framework for agricultural collaboration. First, we introduce a novel combination of differentially private algorithms (like Laplace noise addition) and dimensionality reduction techniques (like Principal Component Analysis) to safeguard farmers' sensitive data while preserving its utility for collaborative research. Dimensionality reduction algorithms compress the data from a higher dimension to a lower dimension, which obfuscates the original feature space. At the same time, Laplace Noise Addition is a technique used to achieve differential privacy. A combination of these two techniques provides a robust guarantee of security against malicious attacks like membership inference attacks (see details in Section~\ref{section:eval}). 
Building on this privacy foundation, we then apply clustering algorithms, such as K-Means, to identify and group farmers with similar characteristics, enabling the formation of networks for meaningful collaboration. By grouping farmers based on shared traits, such as crop types, soil composition, or farming practices, the framework facilitates knowledge exchange, the sharing of best practices, and collective problem-solving. This allows farmers to connect with similar farmers based on the similarity of their farm profiles.  This creates a collaborative environment that not only fosters peer learning but also empowers farmers to adopt innovative techniques, enhance productivity, and address common challenges through resource sharing and mutual support. 

The proposed framework also achieves a robust balance between privacy preservation and utility, enabling effective training of ML models. It ensures that the transformed privacy-protected data aggregated by the sandbox from the farmers preserves the utility required for the effective training of models. Evaluation of the framework on real-world datasets like the Wisconsin Farmer's Market and Crop Recommendation dataset \cite{cropDataset} indicates that models trained using the proposed architecture achieve comparable accuracies to those trained on centralized raw data (see details in Section~\ref{sec:CaseStudy}).

Our framework combines privacy preservation with a sandboxed collaboration environment, enabling researchers to derive insights without exposing raw farmer or market data and providing a novel solution for collaborative research in the agricultural domain, where such systems are absent despite widespread demand.  We envision it to be deployed as a web-based sandbox environment housed on a cloud computing platform like AWS or Azure, featuring a user-friendly interface and an automated backend for data integration, data processing, model training, and sharing. The design ensures easy adoption by farmers and researchers without requiring technical expertise. 

The remainder of this work is structured as follows. To lay the groundwork for our proposed solutions, we start with a literature review of foundational concepts and related works in \textbf{Section \ref{sec:background}: Background and Related Work }. Next, \textbf{Section \ref{section:solution}: Proposed Solution} delves into the design and technical details of our framework. Then, we examine the privacy and utility evaluation metrics and performance of our framework in \textbf{Section \ref{section:eval}: Privacy and Evaluation}. \textbf{Section \ref{section:applications}: Applications in Machine Learning} explores the application of our framework to real-world scenarios and demonstrates the practical utility of our approach. Finally, \textbf{Section \ref{section:conclusion}: Conclusion} summarizes our contributions and discusses their potential impact.

\section{Background and Related work}
\label{sec:background}

This section briefly presents literature relevant to our study, including data privacy challenges in digital agriculture, existing frameworks and mechanisms for secure data sharing, and data privacy techniques. Each provides insights into the current state of research and the gaps our work aims to address.

\subsection{Agricultural Privacy Concerns}
As the agricultural industry evolves, farmers are increasingly adopting technologies to manage soil, water and harvest more efficiently. Since 2010, these innovations, collectively known as Climate-Smart Agriculture or precision agriculture, have transformed large-scale farming practices by enabling real-time monitoring and input optimization \cite{FAO, lowenberg2019setting}. These practices are critical for commodity farmers, improving resource use efficiency and sustainability by tailoring inputs to specific crop conditions. However, data-driven technologies also risk deepening existing inequalities in the agricultural sector \cite{taylor2018climate}, particularly among small and diversified farmers.

Small and medium-sized farmers face significant barriers to adopting data-driven technologies. First, these tools are predominantly designed for large-scale, monoculture farming systems, making them less applicable to diversified farms with varying crops and conditions \cite{WAKWEYA2023100698, klerkx2019review}. Second, while large-scale farmers often share data with industry partners to receive tailored recommendations, small-scale farmers are hesitant due to concerns over potential data misuse \cite{taylor2018climate, WAKWEYA2023100698, WESTERMANN2018283}. Since 2015, academic research has increasingly recognized these privacy concerns, proposing legislative protections and software solutions to alleviate these challenges \cite{linsner2021role, Kaur2022, wolfert2017big}. Despite these efforts, as of August 2024, the agricultural sector lacks a dedicated federal regulatory body for data protection—unlike the healthcare (HIPAA) and finance (GBLA) sectors \cite{ferris2017data}. While the U.S. Cybersecurity and Privacy Operations Center (CPOC) oversees some agricultural data concerns \cite{USDA}, it lacks the authority to enforce industry-wide protections \cite{ferris2017data}.

Agricultural data storage involves multiple stakeholders, including Agricultural Technology Providers (ATPs), farmers, agronomists, and digital platforms. Emerging technologies, such as edge computing and blockchain, promise robust data security and traceability \cite{amiri2022big, kamilaris2019rise}. However, small-scale farmers face substantial challenges in adopting these advanced systems, often bearing significant risks without receiving proportional benefits \cite{wiseman2019farmers, wolfert2017big}. Additionally, the complexity of lengthy, technical data-sharing agreements further discourages smallholder participation, exacerbating their concerns about data misuse and undermining trust in digital agriculture systems \cite{kosior2018digital}.

\subsection{Data Sharing Frameworks for Digital Agriculture}
The literature on data-sharing frameworks in agriculture has addressed privacy and security concerns. For instance, \cite{SPANAKI2021102350} propose a set of principles and an access control mechanism within Data Sharing Agreements (DSAs) to regulate access to shared data and address privacy concerns in the agricultural sector. Similarly, \cite{9170612} introduces a flexible, cloud-based, privacy-preserving data aggregation scheme that allows users to access aggregated plain-text data while safeguarding farmers’ sensitive information. These frameworks offer scalable solutions for secure data sharing; however, their applicability to small and medium-sized farming operations remains limited.

Most studies focus heavily on securing smart devices from external attacks, often overlooking broader privacy frameworks tailored to diversified farming operations. For example,  \cite{Gupta2020} propose a multi-layered architecture that combines cyber-physical measures to deter cyberattacks on smart farming systems. \cite{Kumar2022} extend this work by introducing a privacy encoding framework for sensor-based systems, complemented by an intrusion detection system that maintains data integrity. Similarly, \cite{KUMAR2021107819} developed a privacy-preserving framework for unmanned aerial vehicles (UAVs), emphasizing authentication mechanisms to prevent data poisoning attacks that could compromise agricultural decision-making processes.

Despite these advancements, there is a growing necessity for tailored security frameworks that address the unique challenges faced by small and medium-sized farms, particularly those with diverse farming practices \cite{ongadi2024comprehensive}. To the best of our knowledge, few existing frameworks enable effective and privacy-preserving data collection from farmers, particularly for use in AI and machine learning (ML) applications. Our study addresses this critical gap by proposing and testing a privacy-preserving framework that ensures the confidentiality of individual farmers' data while optimizing the utility of aggregated datasets. By balancing privacy and data usability, our framework empowers researchers to develop AI-driven solutions that are both ethical and impactful for agricultural research.

\subsection{Relevant Data Privacy Techniques}
In the broader context of data-driven approaches, protecting sensitive information is crucial across various domains, including agriculture. While some data privacy techniques have been widely applied in fields such as finance and healthcare, they also hold potential for enhancing data security in digital agricultural applications. This section examines several relevant key data privacy techniques, exploring both their theoretical foundations and their adaptability for use in agriculture. We will discuss Principal Component Analysis (PCA), which, though not explicitly designed for privacy, offers a form of data abstraction by compressing the data into a lower-dimensional space, thereby obfuscating the original feature space; and Local Differential Privacy (LDP), a method that ensures robust privacy through local data perturbation. Understanding the mathematical principles behind these techniques is essential to appreciating their role in safeguarding sensitive data, particularly as the proposed framework is presented.

\textit{\textbf{Principal Component Analysis (PCA)}} \cite{Pearson} is a widely utilized technique for data analysis and the identification of patterns and relationships within data. It has a wide variety of applications, such as image compression \cite{Gaidhane}; facial recognition \cite{Gottumukkal}; medical data correlation \cite{Qureshi}; and Quantitative Finance \cite{Yu}. PCA itself does not inherently provide any guarantees of privacy, as its primary function is to reduce the dimensionality of data by transforming it into a set of orthogonal components. However, this transformation abstracts the data, making it harder to reverse engineer individual records and thereby providing a form of data obfuscation. Similar techniques have been used in the identification of population stratification for genomics research \cite{Dervishi}.

\textit{\textbf{Local Differential Privacy (LDP)}} \cite{Duchi, Kairouz} is a privacy-preserving technique that builds upon the principles of classic Differential Privacy (DP) \cite{Dwork}. LDP is designed to protect individual data and extends the concept of DP to a decentralized context. In LDP, each user perturbs their own data prior to sending it to a data collector. This local perturbation ensures that the privacy of individual data is protected, even in cases where the data collector is not considered reliable. Unlike DP, which relies on a central trusted party that has access to the raw data, LDP offers privacy assurances directly at the data source. An algorithm \( A \) is considered to satisfy \( \epsilon \)-LDP if, for any pair of individual private data points \( a_1 \) and \( a_2 \), and for any possible output \( b \),:

\begin{equation}
\label{equation:1}
\Pr[A(a_1) = b] \leq e^\epsilon \cdot \Pr[A(a_2) = b],
\end{equation}

where \( \epsilon \) represents the privacy parameter. \( \epsilon \)-LDP is achieved by introducing an appropriate amount of noise to individual data points. The primary challenge is determining the noise's magnitude to ensure compliance with LDP while preserving data utility. Various mechanisms have been developed in the DP field to address this problem, with the Laplacian mechanism \cite{Dwork2006} being one of the most prominent.

For any numerical function \( f(x): \mathbb{R} \rightarrow \mathbb{R} \) that satisfies \( \epsilon \) -LDP if Laplacian noise is added as follows:

\begin{equation}
F(x) = f(x) + Lap\left(\frac{s}{\epsilon}\right)
\end{equation}

where \( s \) represents the sensitivity of the function \( f \) and \( \text{Lap}(k) \), where \( k = \frac{s}{\epsilon} \), denotes sampling from a Laplace distribution with scale \( k \). Sensitivity measures the maximum change that a single data point can cause to the output of \( f \) in the worst-case scenario. In our work, we calculate sensitivity to add Laplacian noise \cite{Dwork2013}. By definition, \( s = \max_{x, x'} \|f(x) - f(x')\|_1 \).


\section{Proposed Solution} 
\label{section:solution}

\subsection{System Model}
Our proposed privacy-preserving comprises two main actors: researchers $R$ and participants $P_i$ (e.g., farmers). Participants generate, own, and retain their data to derive insights and benefits from it. Researchers, on the other hand, need the data to conduct research analysis and machine learning model training (see Table~\ref{table:symbols}). The relationship between researchers and participants is inherently symbiotic as researchers need data to do their work and ultimately, their work provides benefits to the farmers in the form of machine learning models such as predictive models for disease detection or weather forecasting. The framework is a sandbox environment that performs key operations, such as data aggregation, clustering, and filtering in the middle. Together, these components form the complete the entire architecture.

\subsection{Threat Model}
In the proposed framework we assume the farmers to be trusted entities as they provide data to the researcher. They generate private agricultural data and perform local PCA transformations and noise addition (to achieve differential privacy for the shared data) before sharing the transformed data. 

We assume the researcher to be an honest but curious party. They  follow the protocol's specifications correctly. However, they might be curious to learn unauthorized information about farmers' private data through protocol outputs, which include analyzing the transformed data matrices, clustering results, and machine learning models. The researcher has access to public agricultural datasets and can perform polynomial-time computations on all protocol outputs. 

We also assume the server to be honest but curious. Nonetheless,
the protocol protects against data leakage to the server by implementing differential privacy techniques. These techniques add noise to the data, obscuring individual farmers' sensitive information. The protocol also utilizes dimensionality reduction methods, such as PCA, to further obfuscate the data structure, making it challenging for attackers to extract meaningful insights from the interactions they view.

We consider various adversarial attacks like membership inference attacks, attribute inference attacks, and deanonymization attacks. Since the identities of the participants who share the data are hidden, attribute inference attacks are only feasible if an attacker can infer the victim's membership to the research dataset through a membership inference attack. The proposed protocol is designed to be robust against membership inference attacks through several mechanisms. We evaluate the robustness of our framework against inference attacks by performing power analysis (see Section~\ref{section:eval}). Although deanonymization attacks can be effective when auxiliary data is available, obscuring original feature space using dimensionality reduction and applying noise to shared datasets significantly reduces its success. 

Each farmer independently performs data transformation and noise addition, ensuring that compromising one farmer's process cannot affect others. The framework infrastructure enforces a strict separation between different farmers' data streams, preventing unauthorized mixing or correlation of data from various sources. The framework infrastructure is assumed to maintain its integrity and not collude with any party to breach privacy guarantees.

\begin{table}
\centering
\caption{Frequently used notations.}
\label{table:symbols}
\begin{tabular}{ll}
\hline
\textbf{Notation} & \textbf{Description} \\
\hline
$R$ & Researcher \\
$P_i$ & Participant $i$ \\
$D_s$ & Dataset used by the researcher to train the PCA model \\
$M_s$ & Trained PCA model \\
$O_i$ & Original PCA matrix of participant $P_i$ \\
$C_i$ & Noisy PCA matrix of participant $P_i$ \\
\hline
\end{tabular}
\end{table}

\subsection{Proposed Scheme}
\begin{figure*}[!t]
    \centering
    \includegraphics[width=0.9\textwidth]{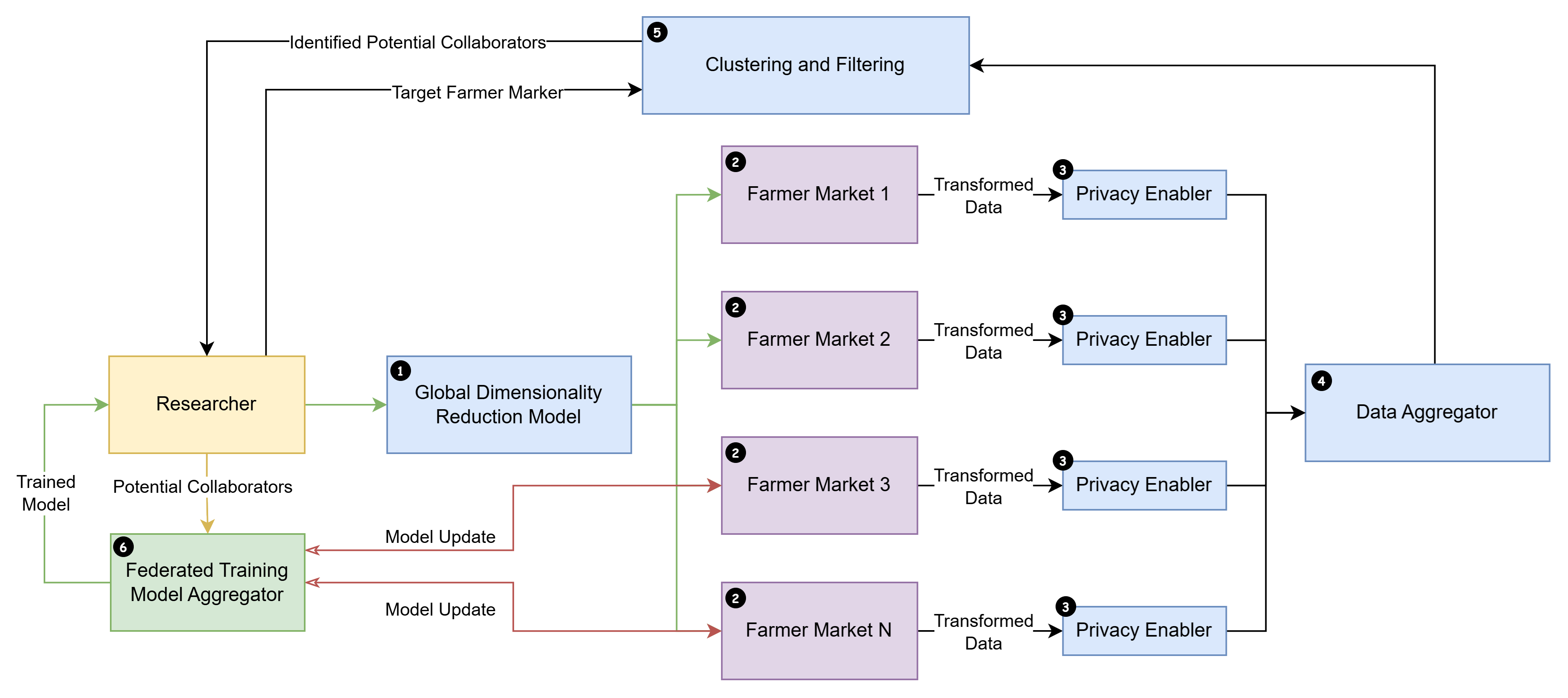}  
    \caption{A high-level data flow: the researcher sends a globally trained dimensionality reduction model to N farmer markets, who transform and anonymize their data before sending it to a data aggregator for sharing on the sandbox platform. The researcher then identifies potential collaborators to train a personalized model.}
    \label{fig:figure1}
\end{figure*}

\begin{figure*}[!t]
  \includegraphics[width=0.9\textwidth]{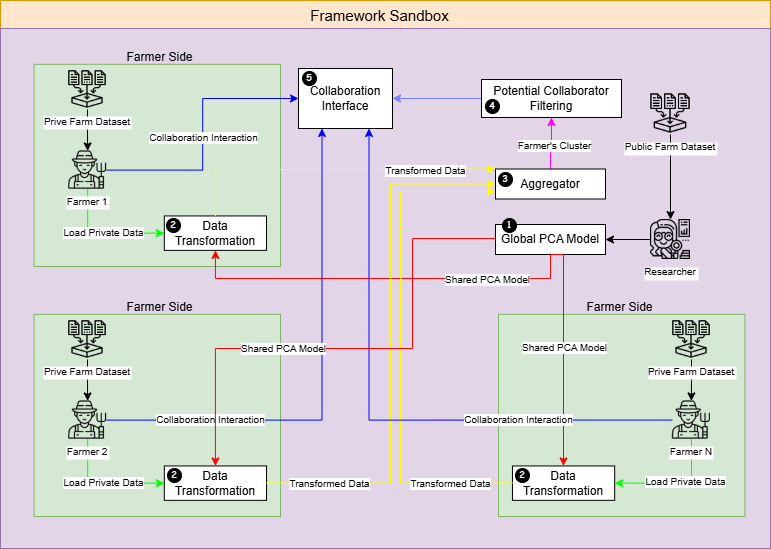}  
  \caption{Detailed architecture of the proposed framework.}
  \label{fig:figure2}
\end{figure*}

We propose a privacy-preserving mechanism using dimensionality reduction and differential privacy techniques to identify farmers having a specific set of attributes. Figures \ref{fig:figure1} and \ref{fig:figure2} illustrate the framework design and workflow. Figure \ref{fig:figure1} illustrates a high-level conceptual overview of the framework's architecture and workflow, free from technical distractions. It provides a high-level system figure, and it emphasizes the end-to-end process, focusing on roles, data flow, and anonymization at an abstract level to enhance clarity for the audience. Figure \ref{fig:figure2} presents a more detailed overview of the proposed framework, concentrating on the granular technical aspects of the core components, data flow, data processing, and infrastructure details essential for replicability and a deeper understanding. 
See Table~\ref{table:symbols} for the list of notations and their definitions. 

The researcher $R$ trains a global dimensionality reduction model such as a principal component analysis (PCA) model on a publicly available dataset $D_s$ having desired farm features (e.g., nitrogen levels, annual rainfall, or pH). The global model is generated to ensure that participants can map their locally reduced data into a common space, allowing for consistent and comparable results when the data is integrated or analyzed together. We use the Standard Scaler function from scikit-learn \cite{pedregosa} to preprocess dataset $D_s$. This trained dimensionality reduction model $M_s$ is then shared with all the participants through the sandbox $F$ along with the sensitivity parameter, which will be used later in the noise addition step. 
Each $P_i$ uses model $M_s$ to transform its private data into transformed matrix $O_i$. To guarantee the privacy of the data, especially against membership inference attacks, the framework applies privacy enhancement techniques like adding Laplacian noise locally to achieve $\epsilon$-LDP. Noise is added to each component of a $O_i$ matrix using \ref{equation:1}, where sensitivity $s$ is replaced with $s_i$, which denotes the sensitivity of component $i$ in the $O_i$ matrix. Finally, a differentially private version of their transformed data $C_i$  is sent to sandbox $F$. 

The sandbox environment receives differentially private data collected from multiple farmers, aggregates it, and then applies clustering algorithms to group farmers.  Figure \ref{fig:figure2} presents the complete workflow of the framework, with a specific use case of connecting similar farmers based on farm attributes. Farmers interact with the sandbox by uploading their farm profile, which is then input to a clustering algorithm, such as K-Means \cite{Hartigan}, pre-trained on aggregated data. The clustering labels are utilized to group farmers and recommend connections among those with similar characteristics. Any $P_i$ (farmer) can query the sandbox with their farm profile to identify potential collaborators using aggregated data. Farmers can communicate, share knowledge, and collaborate with recommended potential collaborators. 

Similarly, the researcher $R$ can utilize the aggregated data to query the sandbox with a target farm profile (e.g., similar to a farm $P_i$) to identify potential collaborators. This helps the researcher $R$ to train various machine learning (ML) models on identified farmers' data. $R$ identifies potential collaborators of a target $P_i$ and uses Federated Learning (FL) techniques to train personalized machine learning models tailored to target $P_i$ profile trained on data from identified potential collaborators. Furthermore, machine learning models can be trained directly on the aggregated data to make predictions. These models can predict various outcomes, such as the likelihood of farm success or failure, or analyze risks like disease outbreaks. It is important to emphasize that $R$ does not have direct access to the private datasets. All analyses and predictions are conducted exclusively using differentially private outputs collected from farmers, ensuring a completely privacy-preserving approach. 


\subsection{Case Study}
\label{sec:CaseStudy}
For feasibility testing and evaluation\footnote{All implementation codes are available on GitHub and will be provided upon request.}, the following case study explores the potential of the proposed privacy-preserving framework for connecting farmers based on attributes and evaluates it using two distinct datasets: a Wisconsin  Farmer’s Market dataset (Table \ref{table:table2}) and a crop recommendation dataset~\cite{cropDataset} (Table \ref{table:table3}). We consider the following use case from the domain of digital agriculture research: The farmers seek to identify potential collaborators based on farm characteristics to collaborate.

\begin{figure}[!t]
    \centering
    \includegraphics[width=\linewidth]{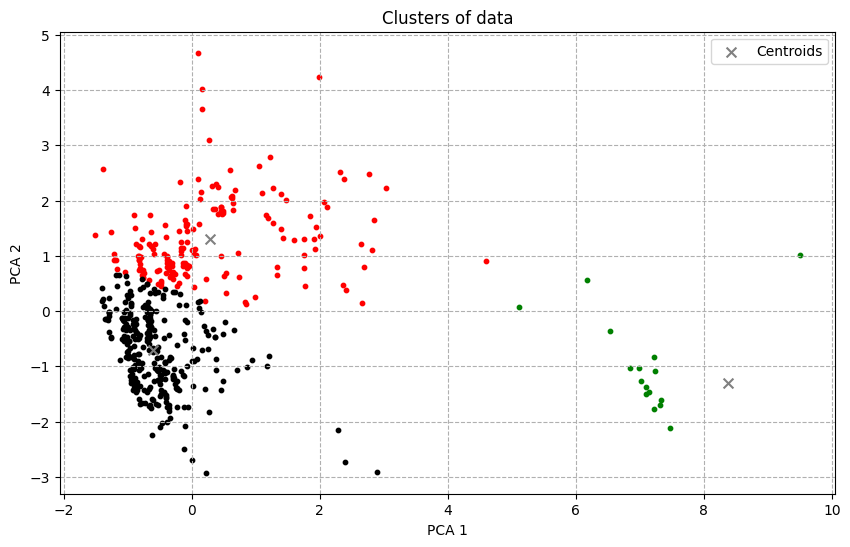}  
    \caption{PCA transformed farmer's market dataset clustering.}
    \label{fig:figure3}
\end{figure}

\begin{figure}[!t]
    \centering
    \includegraphics[width=\linewidth]{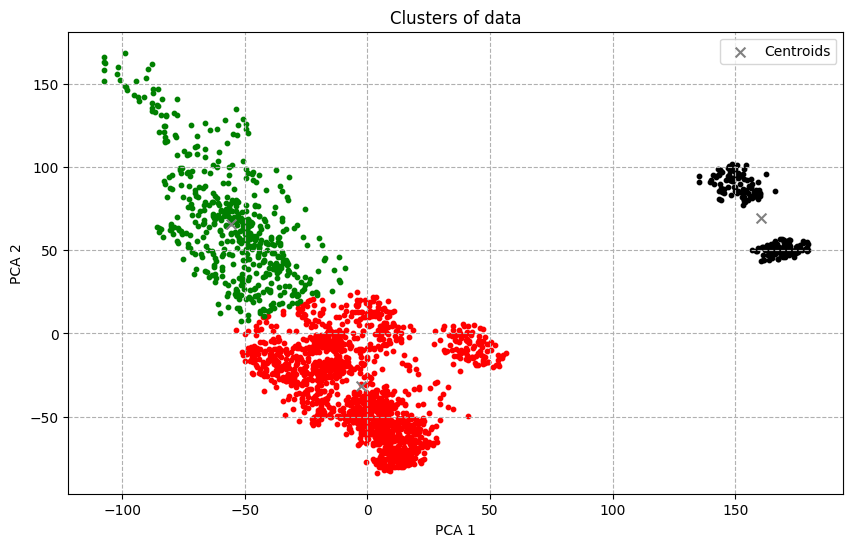}  
    \caption{PCA transformed crop recommendation dataset clustering.}
    \label{fig:figure4}
\end{figure}



\begin{table*}[!t]
\centering
\caption{Farmers market dataset.}\label{table:table2}
\begin{tabular}{cc c c c c c c c c c rr}
\toprule
    & \textbf{Miles from Market} 
    & \multicolumn{9}{c}{\textbf{Vendor Type}}
    & \textbf{} \
    & \textbf{} \\ 
\cmidrule(lr){3-11}
    \textbf{ID} 
        & \textbf{Primary Location} 
        & \textbf{F\&V} 
        & \textbf{M\&S} 
        & \textbf{D} 
        & \textbf{Eggs} 
        & \textbf{P\&F} 
        & \textbf{N\&L} 
        & \textbf{VA} 
        & \textbf{PF} 
        & \textbf{CAS} 
        & \textbf{Sales} 
        & \textbf{\#Visitors} \\ 
\midrule
6815 & 42.22 & 0 & 0 & 1 & 0 & 0 & 0 & 1 & 1 & 0 & 414.0 & 70 \\
5991 & 19.62 & 1 & 1 & 0 & 0 & 0 & 0 & 0 & 0 & 0 & 157.0 & 404 \\ 
5663 & 23.54 & 0 & 0 & 0 & 0 & 0 & 0 & 0 & 0 & 1 & 670.0 & 6 \\ 
5950 & 70.52 & 0 & 0 & 0 & 0 & 0 & 0 & 1 & 0 & 0 & 70.0 & 53 \\ 
5686 & \phantom{0}0.68 & 0 & 0 & 0 & 0 & 0 & 0 & 0 & 0 & 0 & 285.0 & 166 \\
\bottomrule
\end{tabular} \\
{\footnotesize F\&V: Fruits and Vegetables; M\&S Meat and Seafood; D:Dairy; P\&F: Plants and Flowers; N\&L: Nuts and Legumes;\\ VA: Value-added; PF: Prepared Food; CAS: Crafts/Art/Services.}
\end{table*}

\begin{table*}[!t]
\centering
\caption{Crop recommendation dataset.
}\label{table:table3}
\begin{tabular}{cc c c c c c c c c c rr}
\toprule

    \textbf{N} 
        & \textbf{P} 
        & \textbf{K} 
        & \textbf{Temp} 
        & \textbf{H} 
        & \textbf{pH} 
        & \textbf{R} 
        & \textbf{Label}  \\ 
\midrule
90 & 42 & 43 & 20.879 & 82.002 & 6.502 & 202.935 & rice \\
61 & 44 & 17 & 26.100 & 71.574 & 6.931 & 102.266 & maize \\
40 & 72 & 77 & 17.024 & 16.988 & 7.485 & 88.551 & chickpea \\
13 & 60 & 25 & 17.1636 & 20.595 & 5.685 & 128.256 & kidney beans \\
90 & 46 & 42 & 23.978 & 81.450 & 7.502 & 250.082 & rice \\
\bottomrule
\end{tabular} \\
{\footnotesize N: Nitrogen; P: Phosphorous; K:Potassium; pH: pH value; R: rainfall in mm; \\Temp: temperature in Celsius; H: relative humidity in \%.}
\end{table*}

Farmer’s Market dataset (Table \ref{table:table2}), which was gathered internally and has not been made public, includes detailed information about various aspects of farmer's market activities and vendors and was explicitly collected for research purposes. Due to the private nature of the information contained in this dataset, it remains private. Meanwhile, the crop recommendation dataset (Table \ref{table:table3}) is a publicly available dataset collected from multiple farms representing the soil composition and environmental factors~\cite{cropDataset}. The dataset is divided into five pieces representing various markets with uneven sample distributions for each crop, collecting data from different farmers and sharing it with researchers. We consider a researcher receiving the dataset in a privacy-preserving manner using the aforementioned techniques and applying a clustering algorithm like K-Means. Figure \ref{fig:figure3} and Figure \ref{fig:figure4} illustrate the plot of K-Means clustering on the PCA-transformed farmer's market dataset in Table \ref{table:table2} and crop recommendation dataset in Table \ref{table:table3}, respectively. While recommending potential collaborators, the farm profile is fed to the K-Means model to get the group label for the farm. Then, we use the nearest neighbor algorithm \cite{nearestNeighbor} to find the nearest neighbors of the farm profile having the same group label.

\begin{figure}[!t]
    \centering
    \includegraphics[width=\linewidth]{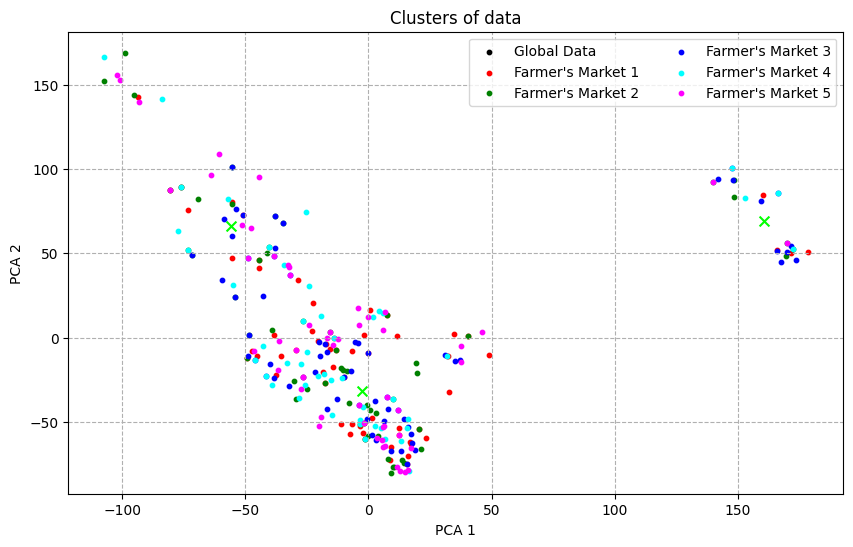}  
    \caption{K-Means clustering on the PCA-transformed crop recommendation dataset, grouped into five farmer's markets and a global set.}
    \label{fig:figureMarket}
\end{figure}
\begin{figure}[!t]
    \centering
    \includegraphics[width=\linewidth]{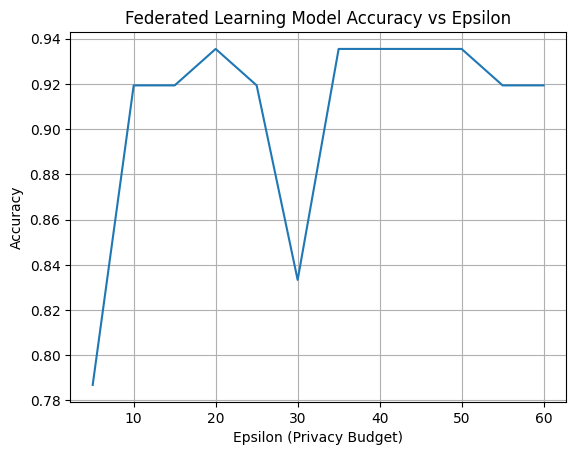}  
    \caption{Federated learning accuracy vs. epsilon (based on profile similarity).}
    \label{fig:figureFedML}
\end{figure}
\begin{figure}[!t]
    \centering
    \includegraphics[width=\linewidth]{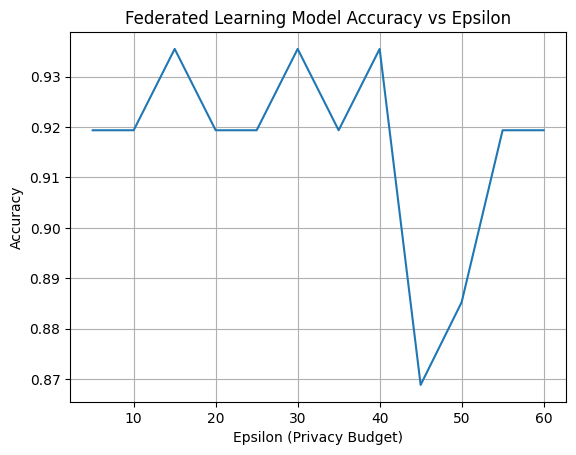}  
    \caption{Federated learning accuracy vs. epsilon (based on data distribution similarity)}
    \label{fig:figureFedML2}
\end{figure}

The entire process is repeated at the Farmer's Market dataset level to identify potential collaborators for personalized machine learning model training via federated learning (see Figure~\ref{fig:figure1}, which illustrates the end-to-end lifecycle of the framework, showcasing one of its use cases: how the researcher uses aggregated data to identify partners and train personalized models.). The crop recommendation dataset (Table~\ref{table:table3}) is divided into a global training dataset and multiple farmer's market datasets to simulate experiments. 
Figure~\ref{fig:figureMarket} illustrates the plot of K-Means clustering on the PCA-transformed crop recommendation dataset in Table \ref{table:table3} divided into five distributed farmer's markets and a global dataset. For personalized model training, a participant $P_i$ (framer's markets in this case) initiates the model training. Frameworks identify three potential collaborators of the initiator $P_i$ out of 5 potential collaborators who can contribute to and participate in training machine learning models under a federated learning setting, enabling secure and collaborative research. There are two ways for a framework to identify potential collaborators: (i) based on the similarity of the participant's profile (data) with each other, and (ii) based on the similarity of distribution of participant's data. Framework uses the nearest neighbor algorithm \cite{nearestNeighbor} to sort all the participants in descending order of their similarity and selects the first three. The trained model is personalized for the initiator participant. Figure \ref{fig:figureFedML} and Figure~\ref{fig:figureFedML2} illustrate the accuracy of the federated learning model vs. the epsilon privacy budget for a neural network trained on three identified collaborators of the initiator $P_i$. In Figure~\ref{fig:figureFedML}, the collaborators are selected based on the similarity of their participant profiles, while in Figure~\ref{fig:figureFedML2}, the collaborators are chosen based on the similarity of the distribution of their data. 

\section{Security Analyses}
\label{security}

The security of the proposed framework reduces to the security of the underlying $\epsilon$-LDP mechanism. In the following, we provide the formal security theorem and its proof.

\newtheorem{theorem}{Theorem}
\begin{theorem}
\label{sectheorem}
Let $M_s$ be the randomization mechanism (e.g., Laplacian mechanism) used in the reduction model. The privacy-preserving scheme for the collaborative agriculture research sharing framework is secure and protects the privacy of the farmers against PPT semi-honest adversaries if the PCA transformation and Laplacian noise satisfies $\epsilon$-Local Differential Privacy.

\end{theorem}

We prove theorem \ref{sectheorem} by introducing two worlds. There is an ideal world the adversary's view \(\text{View}_A^{\text{Ideal}}\) is the outputs of the interactions with a simulator who has no access to the datasets. Moreover, in the real world, the adversary's view \(\text{View}_A^{\text{Real}}\) is the actual output of the framework using the reduction model and Laplacian noise mechanism. The $\epsilon$-LDP guarantee ensures that the outputs of the mechanism for neighboring datasets are statistically close (up to a factor of $e^\epsilon$), meaning that the adversary’s view in the ideal world is computationally indistinguishable from the real world view.

\[
\text{View}_A^{\text{Real}} \overset{c}{\approx} \text{View}_A^{\text{Ideal}}
\]

Where \(\overset{c}{\approx}\) indicates that no PPT adversary \(A\) can distinguish between the two views with non-negligible advantage.

\begin{proof}

In the ideal world, the simulator  $S$ accesses the privacy parameter $\epsilon$ and the adversary’s queries and cannot access the actual datasets $D_s^0$ or $D_s^1$. For each query made by the adversary, $S$ generates noisy outputs using an ideal $\epsilon$-LDP mechanism. The outputs are indistinguishable from those in the real world because of the $\epsilon$-LDP guarantee.

The real world interacts with the adversary $A$ by answering their queries and providing noisy outputs according to the actual interactions of the privacy-preserving framework.

If $A$ has a non-negligible advantage in distinguishing between the two worlds, we can construct a distinguisher $D$ that breaks the $\epsilon$-LDP guarantee of the mechanism $M_s$. 

The distinguisher $D$ receives the output $C_i$ generated from $M_s$ for one of two neighboring datasets $D_s^0$ or $D_s^1$ and simulates the real-world mechanism using $C_i$ as the output of $M_s$. Then, it interacts with the adversary $A$ to send $A$’s queries to $M_s$, generating noisy outputs as needed. $D$ observes $A$’s decision and uses it to distinguish which dataset $C_i$ was generated from. If $A$ can distinguish between the real and ideal worlds with a non-negligible advantage, then $D$ can distinguish between the outputs of $M_s$ applied to $D_s^0$ and $D_s^1$, contradicting the $\epsilon$-LDP guarantee. 

By contradiction, $A$ cannot distinguish between the real and ideal worlds with non-negligible advantages. Therefore, the outputs of the real-world mechanism are computationally indistinguishable from those of the ideal-world mechanism, and the privacy-preserving framework is secure under the $\epsilon$-LDP assumption, which concludes the proof.

\end{proof}

\section{Privacy and Utility Evaluation} 
\label{section:eval}


The farmer data aggregated in a privacy-preserving manner using the proposed framework can be used to train clustering and the nearest neighbor models to find a particular farmer's nearest neighbors (potential collaborators). A researcher can identify potential collaborators and train an ML model on their raw data in a privacy-preserving manner using techniques like Federated Learning. To evaluate the performance of a model trained on data of identified potential collaborators of a particular farmer, we compare its performance with the performance of a model trained on all farmer datasets.

We consider the Wisconsin Farmer's Market dataset (See Table \ref{table:table2}) from our case study section. To simulate two local farmer's market datasets $P_i$s and the dataset used by the researcher to train the PCA model $D_s$, we divide our dataset into three parts. Each $P_i$ converts their data to $C_i$ using shared PCA model $M_s$, which is accessible to the sandbox. 

To evaluate the utility of our process, we compare the accuracies of a classifier trained on raw data aggregated from all the farmers in a centralized manner vs a classifier model trained on data aggregated through the proposed architecture. Table~\ref{table:table4} illustrates the accuracies of various models trained under both settings. The models trained using the proposed architecture exhibit performance comparable to that of the traditionally trained models on centralized data, with an average accuracy loss of 5.33\%. This difference represents a reasonable trade-off, given the substantial privacy benefits achieved.

\begin{table*}[!t]
\centering
\caption{Accuracies of models trained on centralized data vs protected aggregated data at ideal epsilon values.
}\label{table:table4}
\begin{tabular}{cc c c}
\toprule

    \textbf{Model Name} 
        & \textbf{Accuracy on Centralized Data} 
        & \textbf{Accuracy on Aggregated Data} \\ 
\midrule
Logistic Regression & 99.6 & 92.1 \\
Naive Bayes & 98.3 & 93.8 \\
Support Vector Machine & 99.5 & 95.5 \\
\bottomrule
\end{tabular} \\

\end{table*}

\begin{figure}[t]
    \centering
    \includegraphics[width=\linewidth]{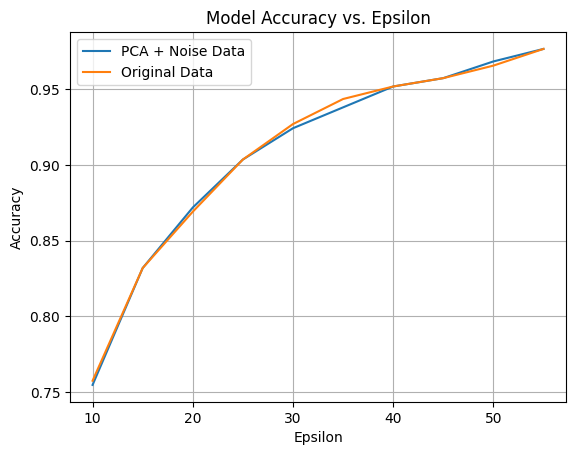}
    \caption{Model accuracy on noisy and original data (Logistic Regression) vs epsilon on crop recommendation dataset.}
    \label{fig:figure11}
\end{figure}

\begin{figure}[t]
    \centering
    \includegraphics[width=\linewidth]{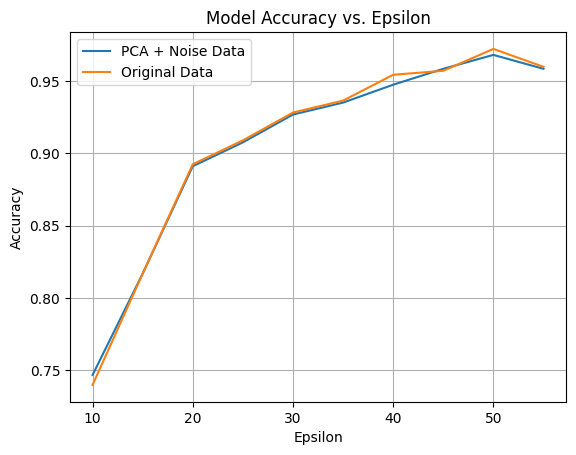}
    \caption{Model accuracy on noisy and original data (Naive Bayes) vs epsilon on crop recommendation dataset.}
    \label{fig:figure12}
\end{figure}

\begin{figure}[t]
    \centering
    \includegraphics[width=\linewidth]{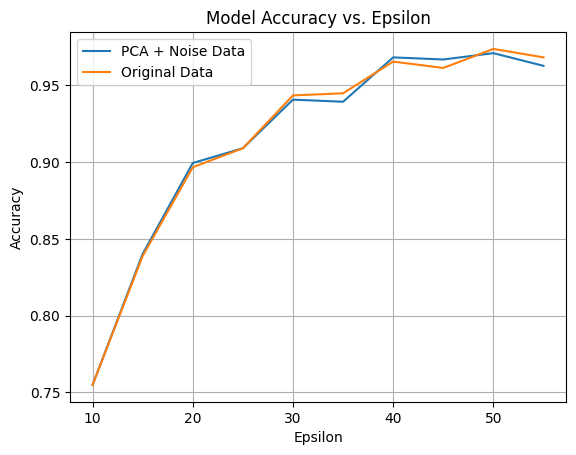}
    \caption{Model accuracy on noisy and original data (SVM) vs epsilon on crop recommendation dataset.}
    \label{fig:figure13}
\end{figure}

To evaluate the effect of noise addition on the utility of the data, we compared model performance on aggregated data with and without noise across varying values of epsilon ($\epsilon$). The parameter $\epsilon$ quantifies privacy protection, where smaller values of $\epsilon$ correspond to stronger privacy guarantees and higher noise, leading to a trade-off with utility. Figures~\ref{fig:figure11}, \ref{fig:figure12}, and \ref{fig:figure13} illustrate the power (fraction of correctly identified noisy samples) and accuracy for a range of $\epsilon$ values for the Logistic Regression, Naive Bayes, and Support Vector Machine (SVM) algorithms, respectively on crop recommendation dataset.  


To evaluate the privacy performance of our algorithm, we compute the power metric, which indicates the fraction of correctly identified noisy samples. To achieve this, we calculate the minimum pairwise distances between the samples of the shared dataset and the control group as the control distances. Similarly, we compute the minimum pairwise distance of the shared dataset and case group as case distances. Here, the control group is sampled from the sandbox global dataset, and the case group is sampled from the participant's shared dataset. Then, we determine the threshold, defined as the distance at the 95th percentile (when the False Positive Rate is 5\%) of the sorted control distances. This threshold is subsequently used to evaluate the fraction of case distances that fall below it, thereby determining the power. The metric is evaluated with specific thresholds for False Positive Rate (FPR) and epsilon ($\epsilon$). $\epsilon$ is a privacy parameter of privacy loss at a differential alteration in the dataset and is allocated across columns according to their sensitivity. For our analysis, we set the FPR at 0.05 (5\%) and incrementally vary the $\epsilon$ value, starting from 10, to observe its impact on the power metric. This approach helps us select the correct value of the $\epsilon$ with acceptable robustness represented by the power metric. 

\begin{figure}[t]
    \centering
    \includegraphics[width=\linewidth]{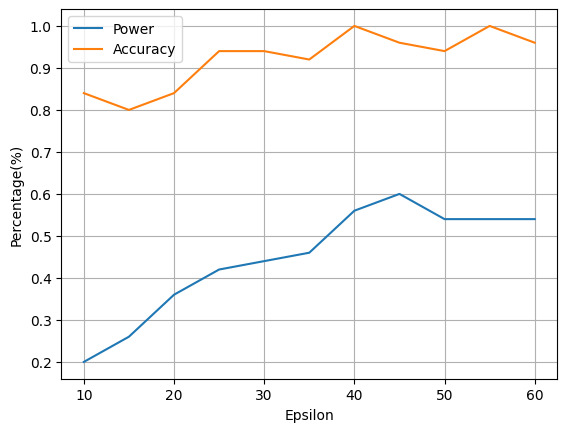}
    \caption{Power and accuracy (Logistic Regression) vs epsilon on crop recommendation dataset.}
    \label{fig:figure5}
\end{figure}

\begin{figure}[t]
    \centering
    \includegraphics[width=\linewidth]{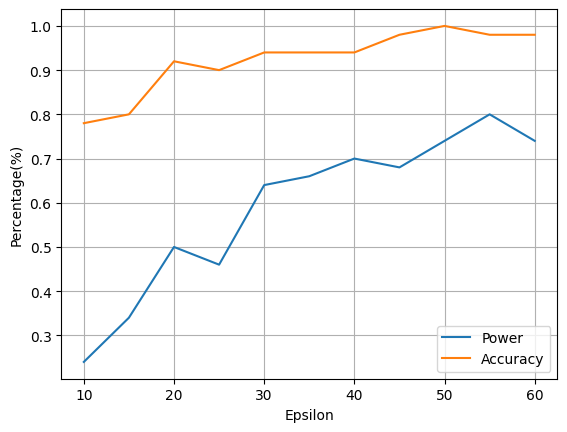}
    \caption{Power and accuracy (Naive Bayes) vs epsilon on crop recommendation dataset.}
    \label{fig:figure6}
\end{figure}

\begin{figure}[t]
    \centering
    \includegraphics[width=\linewidth]{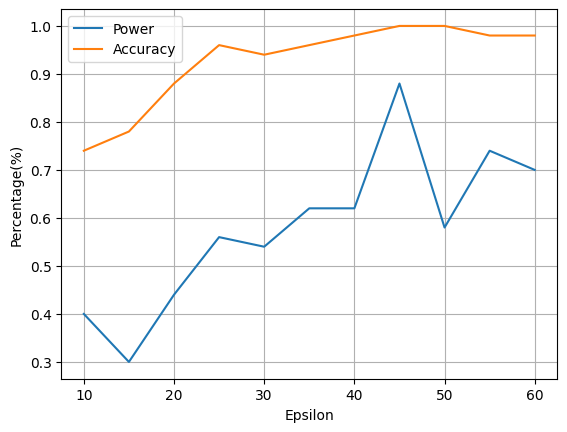}
    \caption{Power and accuracy (SVM) vs epsilon on crop recommendation dataset.}
    \label{fig:figure10}
\end{figure}

\begin{figure}[t]
    \centering
    \includegraphics[width=\linewidth]{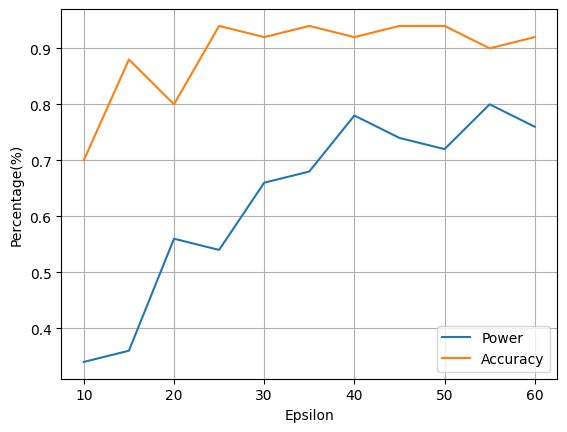}
    \caption{Power and accuracy (Logistic Regression) vs epsilon on farmer's market dataset.}
    \label{fig:figure7}
\end{figure}

\begin{figure}[t]
    \centering
    \includegraphics[width=\linewidth]{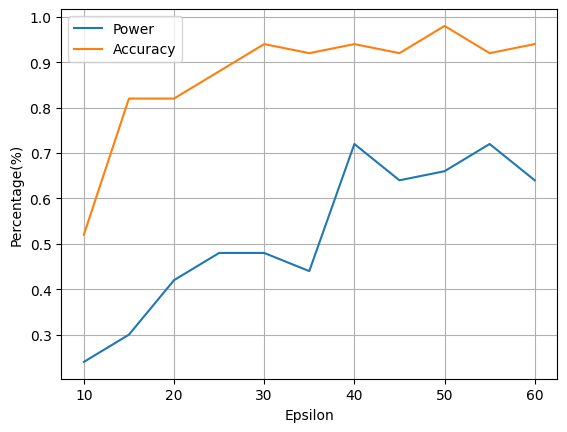}  
    \caption{Power and accuracy (Naive Bayes) vs epsilon on farmer's market dataset.}
    \label{fig:figure8}
\end{figure}

\begin{figure}[t]
    \centering
    \includegraphics[width=\linewidth]{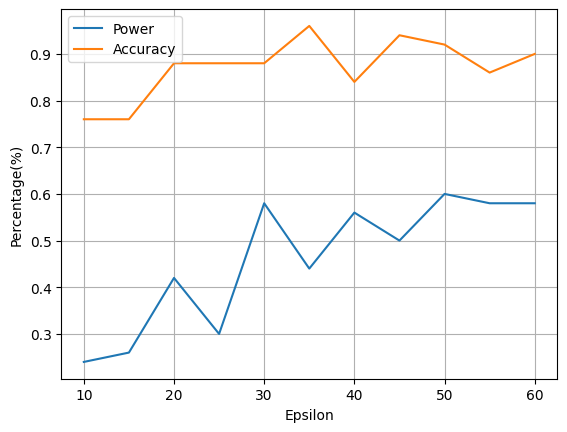}  
    \caption{Power and accuracy (SVM) vs epsilon on farmer's market dataset.}
    \label{fig:figure9}
\end{figure}

For the utility metric, we compare the noise-free cluster labels of the data with the labels predicted by a binary classifier for noisy samples, assessing the accuracy of correctly predicting the labels of noisy samples. We have performed the aforementioned utility test using three different algorithms (Logistic Regression, Naive Bayes, and Support Vector Machine) to demonstrate the results across classifiers. Figures \ref{fig:figure7}, \ref{fig:figure8}, and \ref{fig:figure9} illustrate the power and accuracy for a range of $\epsilon$ values for the Logistic Regression, Naive Bayes, and SVM algorithms, respectively, on the farmer's market dataset. Similarly, Figures \ref{fig:figure5}, \ref{fig:figure6}, and \ref{fig:figure10} illustrate the power and accuracy for a range of $\epsilon$ values for the Logistic Regression, Naive Bayes, and SVM (Support Vector Machine) algorithms, respectively on crop recommendation dataset. The results reflect the high robustness of privacy with comparable utility accuracy. For each model, the optimal $\epsilon$ value is determined based on a trade-off between privacy (measured by the power metric) and utility (measured by accuracy). Specifically, the optimal $\epsilon$ values correspond to those that achieve both low power (indicating strong privacy protection) and high accuracy. Therefore, optimal epsilon values are 25, 35, and 35 for the Logistic Regression, Naive Bayes, and Support Vector Machine models, respectively.


Overall, the framework achieves comparable model accuracies to centralized data training, with minimal utility loss across diverse classifiers (see Table~\ref{table:table4}). This demonstrates that privacy-preserving techniques can be effectively integrated into machine learning models without compromising performance. Operating under realistic assumptions of non-colluding participants and honest-but-curious researchers, a common model in privacy-preserving machine learning, the framework significantly mitigates the effectiveness of adversarial attacks like membership inference attacks, providing a robust defense against privacy breaches. The use of the PCA algorithm transforms data into principle components, thereby obfuscating the original feature space and protecting the original data from reconstruction attacks. 

Additionally, the analysis of noise addition and the power metric further confirm that optimal $\epsilon$ values yield comparable accuracies across various machine learning models, demonstrating the framework’s robustness against adversarial attacks and underscores for real-world collaborative machine learning applications.

\section{Applications in Machine learning} \label{section:applications}
The proposed framework enables researchers to aggregate private data from multiple farmers to train a diverse array of machine-learning models, thereby expanding the scope and quality of their research. We explore a few examples based on the case study dataset that researchers can use to train machine learning models: Researchers can train a model to predict the ideal crop type for a farm based on factors such as soil composition, humidity, and temperature from the market. Additionally, they can utilize farmer's market datasets to develop a model for predicting farm success based on size, distance to market visitors, sales, etc. Alternatively, they can train a machine learning model to predict food accessibility based on factors such as distance from the market, visitor logs, and general public insecurity levels. Similarly, modeling and predicting soil condition data of farms over time to propose management practices and predict crop yields in the region; collecting weather data with soil quality, soil moisture, and crop water requirements to train models for optimizing irrigation systems and predicting water needs in the region; and assessing the impact of climate change on crop yield and soil quality.
\section{Conclusion} \label{section:conclusion}

Privacy concerns remain a significant barrier to sharing sensitive data, particularly when such data could be misused against its owner. This is especially true for farmers, who are often reluctant to share farm data despite its critical role in driving innovation and research in precision agriculture. A robust, privacy-preserving data-sharing solution is, therefore, essential to balance the need for innovation with the protection of sensitive information. This study presents a novel framework that effectively addresses this challenge by employing techniques such as Principal Component Analysis (PCA) and differential privacy. The framework safeguards individual privacy while enabling researchers to leverage aggregated data for machine learning (ML) and AI applications without compromising data utility.

Our contributions advance the field by introducing a privacy-preserving framework that combines dimensionality reduction with differential privacy techniques to enhance both security and data usability. Additionally, the framework facilitates the identification and grouping of farmers with shared characteristics through clustering algorithms, fostering collaboration and collective problem-solving among farmers with similar needs. By addressing these critical gaps, this work provides a scalable and ethical solution for advancing AI-driven research while empowering the agricultural community.

Despite its strengths, the framework has limitations. For instance, the computational requirements of differential privacy techniques may pose challenges for widespread adoption among resource-constrained farming operations. Additionally, the framework’s reliance on consistent and high-quality input data may limit its effectiveness in regions where data collection is inconsistent or less accessible. Future work will focus on refining the framework to enhance its scalability and adaptability to diverse agricultural systems.

In addition, further research will explore integrating decentralized data-sharing mechanisms, such as blockchain, and incorporating real-time data processing to improve trust and responsiveness. By addressing these challenges, this work represents an important step toward enabling privacy-preserving and equitable digital agriculture, laying a foundation for sustainable and ethical farming practices that benefit both researchers and farmers.

\section{Acknowledgments}

This research was supported in part by the National Science Foundation (NSF) under awards OAC-2112606 and 2112533. Also, this research was partly supported by the United States Department of Agriculture (USDA) under grant number NR233A750004G019. 

\newpage
\bibliographystyle{IEEEtran}
\bibliography{sample}

\end{document}